\documentclass[aps,prl,twocolumn,superscriptaddress,showpacs,default] {revtex4-1}
\usepackage{natbib}
\usepackage{color}
\usepackage{graphicx}
\usepackage{dcolumn}
\usepackage{textcomp}
\usepackage{bm}
\usepackage{amsmath}
\begin{document}
\title{Critical current oscillations in the intrinsic hybrid vortex state of SmFeAs(O,F)}

\author{Philip J.W. Moll}
\affiliation{Solid State Physics Laboratory, ETH Zurich, 8093 Zurich, Switzerland}
\author{Luis Balicas}
\affiliation{National High Magnetic Field Laboratory (NHMFL), Tallahassee, Florida 32310, USA}
\author{Xiyu Zhu}
\author{Hai-Hu Wen}
\affiliation{Center for Superconducting Physics and Materials, National Laboratory of Solid State Microstructures and Department of Physics, Nanjing University, Nanjing 210093, China}
\author{Nikolai D. Zhigadlo}
\author{Janusz Karpinski}
\author{Bertram Batlogg}
\affiliation{Solid State Physics Laboratory, ETH Zurich, 8093 Zurich, Switzerland}

\begin{abstract}
In layered superconductors the order parameter may be modulated within the unit cell, leading to non-trivial modifications of the vortex core if the interlayer coherence length $\xi_c(T)$ is comparable to the interlayer distance. In the iron-pnictide SmFeAs(O,F) ($T_c \approx 50$K) this occurs below a cross-over temperature $T^\star \approx 41$K, which separates two regimes of vortices: anisotropic Abrikosov-like at high and Josephson-like at low temperatures. Yet in the transition region around $T^\star$, hybrid vortices between these two characteristics appear. Only in this region around $T^\star$ and for magnetic fields well aligned with the FeAs layers, we observe oscillations of the c-axis critical current $j_c(H)$ periodic in $\frac{1}{\sqrt{H}}$ due to a delicate balance of intervortex forces and interaction with the layered potential. $j_c(H)$ shows pronounced maxima when a hexagonal vortex lattice is commensurate with the crystal structure. The narrow temperature window in which oscillations are observed suggests a significant suppression of the order parameter between the superconducting layers in SmFeAs(O,F), despite its low coherence length anisotropy ($\gamma_\xi \approx 3-5$).
\end{abstract}


\maketitle

Since the discovery of high-temperature superconductivity in the Cu-O planes of the cuprates, the particular dynamics of the vortex matter interacting with layered structures became a central aspect in identifying the microscopic physics as well as in the exploration of the application potential. The vortex shape and supercurrent distribution is influenced by modulations in the surrounding order parameter $\psi$: While for spacially uniform $\psi$ the well-know Abrikosov vortex (AV) with a normal core of diameter $\approx 2\xi$ is formed, full suppression of $\psi$ between the layers leads to Josephson vortices(JV), which lack a normal core but exhibit an extended phase-core region. In between these extreme cases, the situation of partial suppression of $\psi$ may result in the formation of intermediate vortices characterized by deformed core regions. Such intermediate vortices, called hybrid or A-J, have been observed in isolated, strongly coupled junctions such as YBCO low-angle grain boundaries \cite{Gurevich-PRL-2002} or artificial layered structures\cite{Ustinov-PRB-1993,Silin-jetp-1995}, but so far no indications of intrinsic hybrid vortices within the unit cell have been observed.

\begin{figure}[!h]
\centering
\includegraphics[width =0.3 \textwidth]{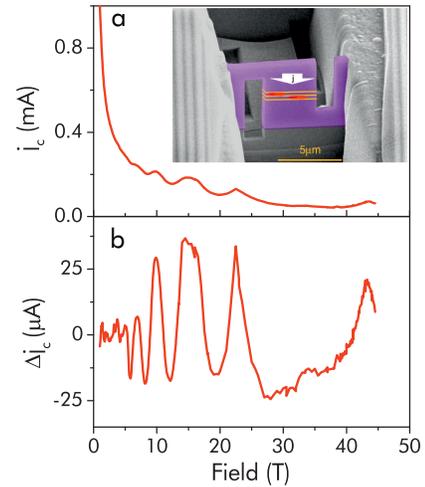}
\caption{ (a) C-axis critical current $j_c(H)$ at $T^\star = 41$K for well aligned in-plane fields, extracted from $I-V$ curves using a fixed voltage criterion. Pronounced oscillations appear ontop of a decreasing background as the field is ramped up. Inset: SmFeAs(O,F) single crystal carved by a FIB into a shape suitable for vortex channel flow between the layers. The active part (purple) is a $3\mu$m long slab along the c-axis ($3$x$1 \mu$m$^2$ in cross-section). The overlaid sketch illustrates the experimental situation of in-plane vortices driven by transverse currents to slide between the FeAs layers. (b) Oscillatory component $\Delta j_c(H)$ after subtracting a power law background. The oscillations first appear around $4$T, and their amplitude grows in field.}
\label{fig:1}
\end{figure}

The more recent advent of the iron-pnictide families as the newest class of high $T_c$ materials places the apparent importance of layeredness and low-dimensionality again into focus. In particular in SmFeAs(O,F) with the highest $T_c^{max} \approx 55$K among the iron-based superconductors, the layered nature determines the microscopic structure of the order parameter: JV centered in the regions of suppressed superfluid density in the Sm(O,F) sheets exist between the FeAs layers, provided that the c-axis coherence length $\xi_c(T)$ is smaller than 1/2 of the c-axis unit cell spacing $d = 0.847$nm. As $\xi_c(T)$ diverges at $T_c$ and shrinks below $d/2$ at low temperatures, a transition temperature $T^\star \approx 41-42$K at $\xi_c(T^\star) \approx d/2$ separates two different regimes: highly mobile JV at low temperatures and well-pinned AV at elevated temperatures\cite{Moll2013}. However, even at zero temperature, $\xi_c(0$K$) \approx 0.18$nm\cite{Welp-PRB-2011} remains comparable to $d/2=0.423$nm, leading to significant remanent interlayer coupling and thus one might well expect the crystal structure of SmFeAs(O,F) to be a candidate to host intrinsic hybrid vortices within the unit cell.

To observe such intrinsic hybrid vortices and their interactions in transport experiments, it is essential to avoid well-pinned Abrikosov-like ``pancake'' segments of the flux lines. Therefore we focus on the ``channel flow'' geometry, in which well aligned in-plane vortices ($<0.1^{\circ}$ in misalignment) slide in between adjacent FeAs planes driven by the Lorentz force of currents along the c-axis. Such a channel flow geometry had been successfully realized in Bi-Sr-Ca-Cu-O mesa structures\cite{Oii-PRL-2002} and recently in Focused Ion Beam (FIB) microcut crystals of SmFeAs(O,F)\cite{Moll2013} and in (V$_2$Sr$_4$O$_6$)Fe$_2$As$_2$\cite{Moll-NatPhys-2014}. The small and platelike SmFeAs(O,F) single crystals were contacted and micro-shaped by a FIB into pillar structures, suitable for vortex channeling experiments (shown in Fig.~\ref{fig:1}a). Details of this technique are described elsewhere\cite{Moll2010}.

The central observation of this study is the sudden appearance of an oscillatory component of the critical current $j_c(H)$ only in the hybrid vortex region around $T^\star$, implying a field-modulated change in the vortex mobility. Figure~\ref{fig:1} shows the critical current $j_c(H)$ at a $5\mu$V criterion and its oscillatory component after subtracting a power-law background. The $j_c$ oscillations become visible at fields beyond $4$T and grow in amplitude at higher fields. At fields beyond $20$T, the almost sinusoidal oscillation changes abruptly to triangular cusps. This may hint at a change of intervortex forces at short vortex distances, and is part of ongoing research. However, pronounced signatures of commensurability enhancement in $j_c$ are observed in high fields beyond $40$T, indicating the formation of a lattice even at very high vortex densities.

\begin{figure}[!h]
\centering
\includegraphics[width =0.45 \textwidth]{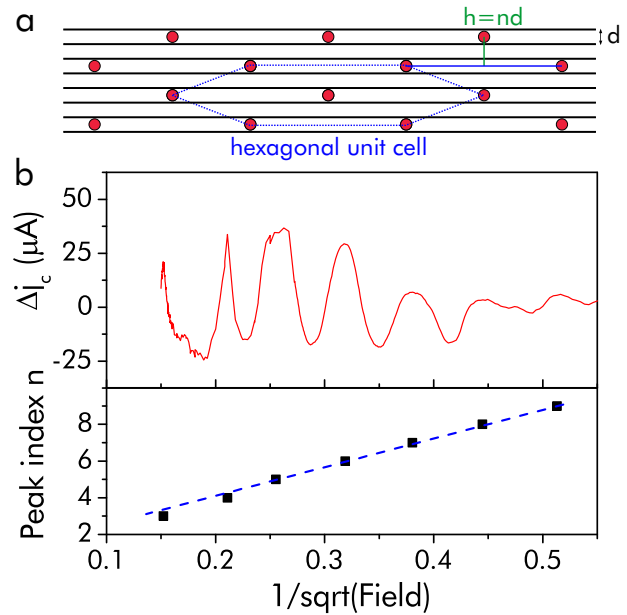}
\caption{(a) Sketch of the hexagonal in-plane hybrid-vortex lattice configuration (red circles indicate core centers) in a matching situation, i.e. with a lattice height $h$ being an integer multiple $n$ of the unit cell spacing $d$. The maxima in $j_c$ are observed when the period of the vortex lattice is commensurate with the underlying lattice. (b) $\Delta j_c$ oscillations and peak indices n as a function of $\frac{1}{\sqrt{H}}$. At large $n$, i.e. low vortex densities, the peak positions are well described by Eq.~\ref{eqn:Hn} (blue dashed line), while the triangular peaks at high fields appear at slightly lower fields.}
\label{fig:2}
\end{figure}

As vortex dynamics generally strongly depends on the local pinning landscape defined by defects in the material, more than 10 samples were studied to distinguish extrinsic sample dependent behavior from intrinsic features generic to SmFeAs(O,F): the oscillations have been observed in all of them consistently\textcolor{red}{(See supplement)}. The maxima of $j_c$ in SmFeAs(O,F) occur at fields denoted by $H_n$, which are equally spaced in $\frac{1}{\sqrt{H}}$ (Fig.~\ref{fig:2}). This $\frac{1}{\sqrt{H}}$ dependence is a natural consequence of a matching between the layered crystal structure and a two-dimensional vortex lattice, which is periodic along and perpendicularly to the FeAs layers. Assuming a hexagonal lattice, deformed by the electronic anisotropy, the values of the matching fields $H_n$ follow from straightforward geometrical considerations. With a field-independent anisotropy parameter $\gamma =  \frac{\sqrt{3}}{2} \frac{a}{h}$, where $a$ denotes the width, $d$ the lattice spacing along c, and $h$ the height of the vortex lattice as indicated in Figure~\ref{fig:2}a, one finds:

\begin{eqnarray}
\Phi_0 = ahH = \gamma \frac{2}{\sqrt{3}} n^2 d^2 H\\
n=\sqrt{ \frac{\sqrt{3}}{2} \frac{\Phi_0}{\gamma d^2}} \frac{1}{\sqrt{H}}
\label{eqn:Hn}
\end{eqnarray}

The only unknown material parameter entering the matching field equation (\ref{eqn:Hn}) is the vortex lattice anisotropy $\gamma$, and thus $\gamma$ can be calculated from the measured slope of $n \propto \frac{1}{\sqrt{H}}$ shown in Fig.~\ref{fig:2}. With $d=0.847$nm for the FeAs interlayer distance\cite{Karpinski-PhysC-2009}, one obtains $\gamma = 9.4$. This value is in excellent agreement with the penetration field anisotropy $\gamma_\lambda(42$K$) \approx 9.6$ around $T^\star$ obtained independently by torque magnetometry\cite{Weyeneth-JSNM-2009,Karpinski-PhysC-2009} and thus strongly indicates the proposed elongated hexagonal vortex lattice to be the appropriate description.

\begin{figure}[!h]
\centering
\includegraphics[width =0.53 \textwidth]{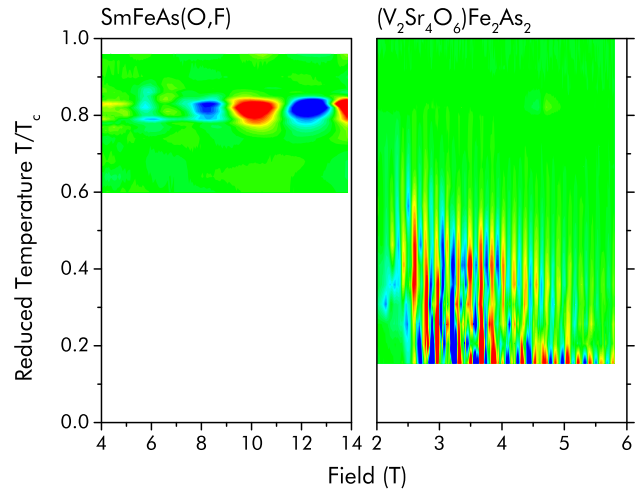}
\caption{Oscillatory component of the critical current $j_c(H)$ after background subtraction. In SmFeAs(O,F) (left), the oscillations exist only in a narrow region around $T^\star$. This is in contrast to other layered superconductors, in which $j_c$ oscillations are observed over a wider temperature range. An example for such a behavior is provided by (V$_2$Sr$_4$O$_6$)Fe$_2$As$_2$ (right). The longer c-axis spacing leads to a stronger Josephson behavior when compared to SmFeAs(O,F), which induces oscillations periodic in field instead of $\frac{1}{\sqrt(H)}$.}
\label{fig:3}
\end{figure}

The above discussion only concerns measurements at the A-J transition temperature $T^\star \approx 0.84 T_c$, while the temperature dependence reveals the essential role of the hybrid vortex nature in the $j_c$ modulations resulting from a subtle balance between Abrikosov-like pinning and Josephson-like channeling. After subtracting a smoothly varying background, the temperature evolution of the oscillatory part of $j_c$ becomes evident: The oscillations in SmFeAs(O,F) are most pronounced at $T^\star$ and are only observable in a narrow temperature window ($\pm 3$K) around $T^\star$ (Fig.~\ref{fig:3} left). There is no indication of any oscillatory component of $j_c$ in the Abrikosov- or in the Josephson-state and thus the vortex interactions leading to this oscillatory phenomenon are a unique property of hybrid vortices in the transition region between Abrikosov and Josephson. This is in contrast to previously observed oscillatory phenomena periodic in $\frac{1}{\sqrt{H}}$ in low anisotropy cuprates, such as YBa$_2$Cu$_3$O$_{7-x}$\cite{Gordeev-PRL-2000,Baily-PRL-2008,Tokita-LowTemp-2009} and NdBa$_2$Cu$_3$O$_x$\cite{Kupfer-PRB-2002}. In these systems, the oscillations also appear at an onset temperature $T_0 < T_c$, however they persist over a wide temperature range down to much lower temperatures. One important difference between these compounds is the multi-band nature in SmFeAs(O,F) compared to the single-band cuprates. The second gap in SmFeAs(O,F) is significantly smaller than the larger gap, and thus multi-band effects are expected to become important at temperatures much below $T^\star$ in agreement with the smooth temperature dependence of $\frac{1}{\lambda^2}$\cite{Malone-PRB-2009} and $H_{c2} \propto \frac{1}{\xi^2}$\cite{Moll2010} around $T^\star$. Therefore we do not expect multi-band effects to significantly influence the distinct temperature dependence of the oscillations in SmFeAs(O,F).

In contrast, Fig.~\ref{fig:3} (right) shows vortex oscillations in (V$_2$Sr$_4$O$_6$)Fe$_2$As$_2$, which due to its larger and more insulating spacing layer and hence significantly larger $V_0$ behaves as a fully developed Josephson system similar to e.g. Bi-Sr-Ca-Cu-O\cite{Moll-NatPhys-2014}. In this case, only a one-dimensional vortex configuration is observed within each layer, leading to oscillations periodic in field (and not in $\frac{1}{\sqrt{H}}$) due to modulations of the surface barrier\cite{Koshelev-PRB-2007,Machida-PRL-2003,Oii-PRL-2002}. These oscillations also appear below an onset temperature $T_0$, however they persist down to the lowest temperatures accessed by the experiment as they do not require vortex mobility along the c-direction.

\begin{figure*}[htb]
\centering
\includegraphics[width =0.85 \textwidth]{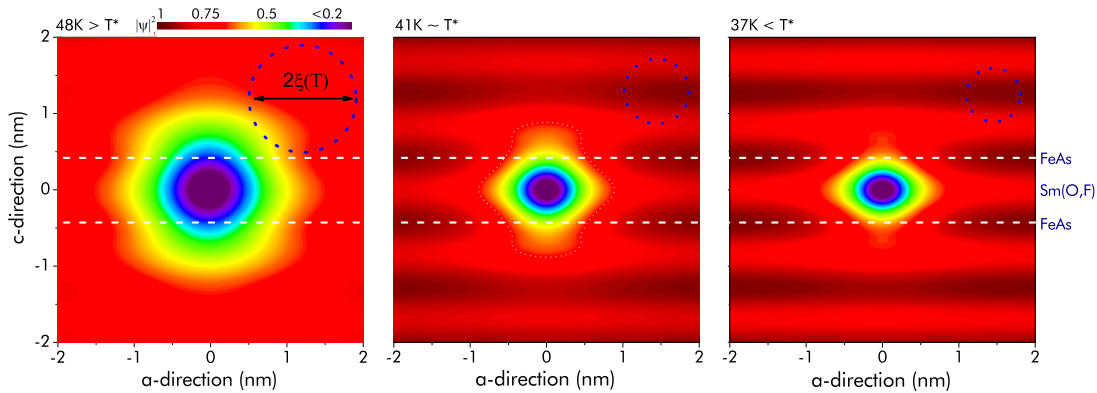}
\caption{Order parameter modulus $\lvert \psi \rvert^2$ in a situation of slightly suppressed OP in the SmO layer following Eq.~\ref{eqn:GL}. (top panel) Large Abrikosov vortex at high T spans several unit cells. (middle) Hybrid vortex at $T^\star$ gains condensation energy by centering the core in the SmO layer of suppressed OP, yet there still is a significant suppression of the OP in the adjacent FeAs layers leading to c-direction mobility. (bottom) Below $T^\star$, the vortex is completely confined between two FeAs layers. The suppression of the superfluid density limits the local critical current perpendicular to the layers, leading to an elongation of the core along the a-direction. }
\label{fig:4}
\end{figure*}

To qualitatively understand the influence of the hybrid vortex nature on the temperature evolution of the oscillations, we have numerically studied vortex cores in layered structures by solving the Ginzburg-Landau equations in the presence of an in-plane magnetic field. The layered structure of SmFeAs(O,F) was modeled by adding a step-function potential energy term $V(z)\lvert \psi \rvert^2$ into the free energy functional (Eq.3), thus partially suppressing the superfluid density in the Sm(O,F) layers. The effective thickness of the superconducting layer was assumed to be $\frac{1}{2} d$, according to the geometric extent of the Sm(O,F) layer in the unit cell determined by X-Ray diffraction\cite{Karpinski-PhysC-2009}. This time-independent problem was then solved numerically\cite{Gropp1996} using the finite element solver COMSOL. To emphasize the appearance of a vortex core anisotropy solely due to its interaction with the modulated order parameter (OP), an isotropic coherence length $\xi_c=\xi_a$ was used for the calculation shown in Fig.~\ref{fig:4}. In such an isotropic case without the presence of suppression layers, circular vortex cores are expected as indicated by blue dashed circles. Any deviation from the circular shape is only due to the layers of suppressed OP. The coherence length anisotropy can easily be introduced into the results following the scaling relations proposed by Blatter et al.\cite{Blatter-RevModPhys-1994}.

\begin{eqnarray}
\begin{split}
\mathcal{L} = a \lvert \psi \rvert^2 + \frac{b}{2} \lvert \psi \rvert^4 + \frac{1}{2 m_s} \lvert (\frac{\hbar}{i} \nabla - \frac{e_s}{c} \textbf{A}) \psi \rvert^2 \\
 + V(z) \lvert \psi \rvert^2
\end{split}
\\
V(z) = \begin{cases} -V_0, & \mbox{in FeAs layer}  \\ +V_0, & \mbox{in Sm(O,F) layer} \end{cases}
\label{eqn:GL}
\end{eqnarray}

The resulting vortex core regions in the presence of layers of suppressed order parameter are shown in Fig.~\ref{fig:4}. At high temperatures in the Abrikosov state ($\xi_c(T) > d_c$), the superconducting condensate cannot follow the potential modulated at length scales below $\xi$, resulting in large, essentially circular Abrikosov-like cores. Thus commensurability oscillations are naturally absent in the Abrikosov state. In the Josephson-state, however, a very different mechanism suppresses the $\frac{1}{\sqrt{H}}$ oscillations of $j_c$: As the JV (phase-)core is pushed in between two adjacent FeAs layers (lower panel of Fig.~\ref{fig:4}), there is a large energy barrier associated with the movement of the core across the FeAs- and into the adjacent SmO-layer. This motion requires the generation of a pancake vortex-antivortex pair and a successive separation of the two (``zipper mechanism'')\cite{Blatter-RevModPhys-1994}, and therefore the JV motion along the c-direction is effectively suppressed. Vortex entry at the surface is not uniform but occurs predominantly at particular nucleation sites of locally reduced surface barrier \cite{Baelus-PRB-2005}. Establishing the two-dimensional order along as well as perpendicular to the FeAs layers that leads to the $\frac{1}{\sqrt{H}}$ commensurability effects requires the irregularly entering vortices to relax into a lattice. This cannot happen in the JV state without sufficient vortex mobility along the c-axis, and thus the oscillations are absent.

Hence, the two key ingredients theoretically expected in hybrid vortices\cite{Gurevich-PRL-2002} are essential to the observation of this phenomenon in SmFeAs(O,F): (1) A core region small enough to gain energy from aligning with the intrinsic potential, and (2) an incomplete Josephson nature to allow vortex lattice relaxation along the c-axis. This occurs exactly at the $T^\star$ transition, as illustrated in Fig~\ref{fig:4} (middle panel). While the main flux of this hybrid vortex is confined between two adjacent FeAs layers (black dashed line), there is still a substantial suppression of the OP in the FeAs layers in the vicinity of the core center (highlighted by white points). This suppression reduces the barrier that impedes vortex motion along c and thereby allows a two-dimensional vortex lattice to form.

The shape of the vortex core, i.e. the supercurrent profile around the core region, is heavily influenced by the microscopic structure of the OP within the unit cell. Depending on the orbitals involved in Cooper pair transport along and perpendicularly to the layers, the OP modulation along $c$ varies in strength and shape. It will be an interesting theoretical challenge to develop a microscopic model based on the Fe- and As-orbitals\cite{Choubey-arxiv-2014} to derive a more realistic form for the potential $V(z)$. Within the simplistic step-function model, the magnitude of $V_0$ cannot be quantitatively estimated from transport experiments. However, some insight may be gained from a comparison with cuprate systems that show $\frac{1}{\sqrt{H}}$ oscillations.
In YBa$_2$Cu$_3$O$_{7-x}$, similar oscillations persist over an extended temperature region ($>30$K)\cite{Oussena1994} and thus indicate a higher vortex mobility along $c$ than in SmFeAs(O,F) due to a weaker suppression of the OP between the CuO planes, and thus a smaller $V_0$. Therefore the vortices in YBa$_2$Cu$_3$O$_{7-x}$ even at low temperatures show more Abrikosov-like behavior, in particular with a higher mobility between the superconducting Cu-O planes. This picture is supported by differences in the pinning of Josephson-like vortices: The absence of enhanced vortex mobility in the ``channel flow'' geometry (in-plane field and out-of-plane currents) in YBa$_2$Cu$_3$O$_{7-x}$ indicates a highly effective pinning for in-plane vortices below $T^\star$\cite{Lundquist-PRB-2001}, in contrast to the highly mobile vortices in SmFeAs(O,F)\cite{Moll2013}. This suggests a stronger suppression of the OP between the superconducting layers in SmFeAs(O,F), i.e. a larger $V_0$. This difference is even more intriguing in the light of their similar coherence length anisotropies ($\gamma_\xi = 6-5$ in YBa$_2$Cu$_3$O$_{7-x}$\cite{Schnelle-AnnPhys-1993}, $5-3$ in SmFeAs(O,F)\cite{Moll2010}).

In summary, we have found evidence for a significant modulation of the OP within the unit cell of SmFeAs(O,F), leading to commensurability effects between the vortex- and the crystal lattice. In particular, the untypical temperature dependence suggests the existence of hybrid vortices, in between Josephson- and Abrikosov character, in a narrow temperature range corresponding to a cross-over region. In other superconductors showing similar oscillations, i.e. YBa$_2$Cu$_3$O$_{7-x}$, the oscillations persist over much larger temperature ranges. This difference between YBa$_2$Cu$_3$O$_{7-x}$ and SmFeAs(O,F) emphasizes an important role for the strength of the suppression of the superconducting OP in between the layers: Next to the coherence length ($\gamma_\xi$)- and the penetration depth ($\gamma_\lambda$) anisotropy, the suppression strength ($V_0$) is a third parameter of importance to describe the vortex matter in layered superconductors. While these anisotropies are typically related, they are in principle independent and thus may lead to distinctly different vortex behavior even in systems of similar $\gamma$, such as YBa$_2$Cu$_3$O$_{7-x}$ and SmFeAs(O,F).\\

\begin{acknowledgments}
We thank Gianni Blatter and Dima Geshkenbein for stimulating discussions. The FIB work was supported by SCOPE-M (ETH). The NHMFL is supported by NSF through NSF-DMR-0084173 and the State of Florida. L.B. is supportedby DOE-BES through award DE-SC0002613. Work in Nanjing was supported by the Ministry of Science and Technology of China (973 Projects: No. 2011CBA00102, No. 2010CB923002).
\end{acknowledgments}

\bibliography{JcOscBib}

\begin{thebibliography}{24}%
\makeatletter
\providecommand \@ifxundefined [1]{%
 \@ifx{#1\undefined}
}%
\providecommand \@ifnum [1]{%
 \ifnum #1\expandafter \@firstoftwo
 \else \expandafter \@secondoftwo
 \fi
}%
\providecommand \@ifx [1]{%
 \ifx #1\expandafter \@firstoftwo
 \else \expandafter \@secondoftwo
 \fi
}%
\providecommand \natexlab [1]{#1}%
\providecommand \enquote  [1]{``#1''}%
\providecommand \bibnamefont  [1]{#1}%
\providecommand \bibfnamefont [1]{#1}%
\providecommand \citenamefont [1]{#1}%
\providecommand \href@noop [0]{\@secondoftwo}%
\providecommand \href [0]{\begingroup \@sanitize@url \@href}%
\providecommand \@href[1]{\@@startlink{#1}\@@href}%
\providecommand \@@href[1]{\endgroup#1\@@endlink}%
\providecommand \@sanitize@url [0]{\catcode `\\12\catcode `\$12\catcode
  `\&12\catcode `\#12\catcode `\^12\catcode `\_12\catcode `\%12\relax}%
\providecommand \@@startlink[1]{}%
\providecommand \@@endlink[0]{}%
\providecommand \url  [0]{\begingroup\@sanitize@url \@url }%
\providecommand \@url [1]{\endgroup\@href {#1}{\urlprefix }}%
\providecommand \urlprefix  [0]{URL }%
\providecommand \Eprint [0]{\href }%
\providecommand \doibase [0]{http://dx.doi.org/}%
\providecommand \selectlanguage [0]{\@gobble}%
\providecommand \bibinfo  [0]{\@secondoftwo}%
\providecommand \bibfield  [0]{\@secondoftwo}%
\providecommand \translation [1]{[#1]}%
\providecommand \BibitemOpen [0]{}%
\providecommand \bibitemStop [0]{}%
\providecommand \bibitemNoStop [0]{.\EOS\space}%
\providecommand \EOS [0]{\spacefactor3000\relax}%
\providecommand \BibitemShut  [1]{\csname bibitem#1\endcsname}%
\let\auto@bib@innerbib\@empty
\bibitem [{\citenamefont {Gurevich}\ \emph {et~al.}(2002)\citenamefont
  {Gurevich}, \citenamefont {Rzchowski}, \citenamefont {Daniels}, \citenamefont
  {Patnaik}, \citenamefont {Hinaus}, \citenamefont {Carillo}, \citenamefont
  {Tafuri},\ and\ \citenamefont {Larbalestier}}]{Gurevich-PRL-2002}%
  \BibitemOpen
  \bibfield  {author} {\bibinfo {author} {\bibfnamefont {A.}~\bibnamefont
  {Gurevich}}, \bibinfo {author} {\bibfnamefont {M.~S.}\ \bibnamefont
  {Rzchowski}}, \bibinfo {author} {\bibfnamefont {G.}~\bibnamefont {Daniels}},
  \bibinfo {author} {\bibfnamefont {S.}~\bibnamefont {Patnaik}}, \bibinfo
  {author} {\bibfnamefont {B.~M.}\ \bibnamefont {Hinaus}}, \bibinfo {author}
  {\bibfnamefont {F.}~\bibnamefont {Carillo}}, \bibinfo {author} {\bibfnamefont
  {F.}~\bibnamefont {Tafuri}}, \ and\ \bibinfo {author} {\bibfnamefont {D.~C.}\
  \bibnamefont {Larbalestier}},\ }\href {\doibase
  10.1103/PhysRevLett.88.097001} {\bibfield  {journal} {\bibinfo  {journal}
  {Phys. Rev. Lett.}\ }\textbf {\bibinfo {volume} {88}},\ \bibinfo {pages}
  {097001} (\bibinfo {year} {2002})}\BibitemShut {NoStop}%
\bibitem [{\citenamefont {Ustinov}\ \emph {et~al.}(1993)\citenamefont
  {Ustinov}, \citenamefont {Kohlstedt}, \citenamefont {Cirillo}, \citenamefont
  {Pedersen}, \citenamefont {Hallmanns},\ and\ \citenamefont
  {Heiden}}]{Ustinov-PRB-1993}%
  \BibitemOpen
  \bibfield  {author} {\bibinfo {author} {\bibfnamefont {A.~V.}\ \bibnamefont
  {Ustinov}}, \bibinfo {author} {\bibfnamefont {H.}~\bibnamefont {Kohlstedt}},
  \bibinfo {author} {\bibfnamefont {M.}~\bibnamefont {Cirillo}}, \bibinfo
  {author} {\bibfnamefont {N.~F.}\ \bibnamefont {Pedersen}}, \bibinfo {author}
  {\bibfnamefont {G.}~\bibnamefont {Hallmanns}}, \ and\ \bibinfo {author}
  {\bibfnamefont {C.}~\bibnamefont {Heiden}},\ }\href {\doibase
  10.1103/PhysRevB.48.10614} {\bibfield  {journal} {\bibinfo  {journal} {Phys.
  Rev. B}\ }\textbf {\bibinfo {volume} {48}},\ \bibinfo {pages} {10614}
  (\bibinfo {year} {1993})}\BibitemShut {NoStop}%
\bibitem [{\citenamefont {Silin}\ and\ \citenamefont
  {Uryupin}(1995)}]{Silin-jetp-1995}%
  \BibitemOpen
  \bibfield  {author} {\bibinfo {author} {\bibfnamefont {V.~P.}\ \bibnamefont
  {Silin}}\ and\ \bibinfo {author} {\bibfnamefont {S.~A.}\ \bibnamefont
  {Uryupin}},\ }\href@noop {} {\bibfield  {journal} {\bibinfo  {journal} {Zh.
  Eksp. Teor. Fiz.}\ }\textbf {\bibinfo {volume} {108}},\ \bibinfo {pages}
  {2163} (\bibinfo {year} {1995})}\BibitemShut {NoStop}%
\bibitem [{\citenamefont {Moll}\ \emph {et~al.}(2013)\citenamefont {Moll},
  \citenamefont {Balicas}, \citenamefont {Geshkenbein}, \citenamefont
  {Blatter}, \citenamefont {Karpinski}, \citenamefont {Zhigadlo},\ and\
  \citenamefont {Batlogg}}]{Moll2013}%
  \BibitemOpen
  \bibfield  {author} {\bibinfo {author} {\bibfnamefont {P.~J.~W.}\
  \bibnamefont {Moll}}, \bibinfo {author} {\bibfnamefont {L.}~\bibnamefont
  {Balicas}}, \bibinfo {author} {\bibfnamefont {V.}~\bibnamefont
  {Geshkenbein}}, \bibinfo {author} {\bibfnamefont {G.}~\bibnamefont
  {Blatter}}, \bibinfo {author} {\bibfnamefont {J.}~\bibnamefont {Karpinski}},
  \bibinfo {author} {\bibfnamefont {N.~D.}\ \bibnamefont {Zhigadlo}}, \ and\
  \bibinfo {author} {\bibfnamefont {B.}~\bibnamefont {Batlogg}},\ }\href
  {\doibase doi:10.1038/nmat3489} {\bibfield  {journal} {\bibinfo  {journal}
  {Nature Materials}\ }\textbf {\bibinfo {volume} {12}},\ \bibinfo {pages}
  {134} (\bibinfo {year} {2013})}\BibitemShut {NoStop}%
\bibitem [{\citenamefont {Welp}\ \emph {et~al.}(2011)\citenamefont {Welp},
  \citenamefont {Chaparro}, \citenamefont {Koshelev}, \citenamefont {Kwok},
  \citenamefont {Rydh}, \citenamefont {Zhigadlo}, \citenamefont {Karpinski},\
  and\ \citenamefont {Weyeneth}}]{Welp-PRB-2011}%
  \BibitemOpen
  \bibfield  {author} {\bibinfo {author} {\bibfnamefont {U.}~\bibnamefont
  {Welp}}, \bibinfo {author} {\bibfnamefont {C.}~\bibnamefont {Chaparro}},
  \bibinfo {author} {\bibfnamefont {A.~E.}\ \bibnamefont {Koshelev}}, \bibinfo
  {author} {\bibfnamefont {W.~K.}\ \bibnamefont {Kwok}}, \bibinfo {author}
  {\bibfnamefont {A.}~\bibnamefont {Rydh}}, \bibinfo {author} {\bibfnamefont
  {N.~D.}\ \bibnamefont {Zhigadlo}}, \bibinfo {author} {\bibfnamefont
  {J.}~\bibnamefont {Karpinski}}, \ and\ \bibinfo {author} {\bibfnamefont
  {S.}~\bibnamefont {Weyeneth}},\ }\href@noop {} {\bibfield  {journal}
  {\bibinfo  {journal} {Phys. Rev. B}\ }\textbf {\bibinfo {volume} {83}},\
  \bibinfo {pages} {100513(R)} (\bibinfo {year} {2011})}\BibitemShut {NoStop}%
\bibitem [{\citenamefont {Ooi}\ \emph {et~al.}(2002)\citenamefont {Ooi},
  \citenamefont {Mochiku},\ and\ \citenamefont {Hirata}}]{Oii-PRL-2002}%
  \BibitemOpen
  \bibfield  {author} {\bibinfo {author} {\bibfnamefont {S.}~\bibnamefont
  {Ooi}}, \bibinfo {author} {\bibfnamefont {T.}~\bibnamefont {Mochiku}}, \ and\
  \bibinfo {author} {\bibfnamefont {K.}~\bibnamefont {Hirata}},\ }\href@noop {}
  {\bibfield  {journal} {\bibinfo  {journal} {Phys. Rev. Lett.}\ }\textbf
  {\bibinfo {volume} {89}},\ \bibinfo {pages} {247002} (\bibinfo {year}
  {2002})}\BibitemShut {NoStop}%
\bibitem [{\citenamefont {Moll}\ \emph {et~al.}(2014)\citenamefont {Moll},
  \citenamefont {Zhu}, \citenamefont {Cheng}, \citenamefont {Wen},\ and\
  \citenamefont {Batlogg}}]{Moll-NatPhys-2014}%
  \BibitemOpen
  \bibfield  {author} {\bibinfo {author} {\bibfnamefont {P.~J.~W.}\
  \bibnamefont {Moll}}, \bibinfo {author} {\bibfnamefont {X.}~\bibnamefont
  {Zhu}}, \bibinfo {author} {\bibfnamefont {P.}~\bibnamefont {Cheng}}, \bibinfo
  {author} {\bibfnamefont {H.-H.}\ \bibnamefont {Wen}}, \ and\ \bibinfo
  {author} {\bibfnamefont {B.}~\bibnamefont {Batlogg}},\ }\href@noop {}
  {\bibfield  {journal} {\bibinfo  {journal} {Nature Physics}\ }\textbf
  {\bibinfo {volume} {10}},\ \bibinfo {pages} {644} (\bibinfo {year}
  {2014})}\BibitemShut {NoStop}%
\bibitem [{\citenamefont {Moll}\ \emph {et~al.}(2010)\citenamefont {Moll},
  \citenamefont {Puzniak}, \citenamefont {Balakirev}, \citenamefont {Rogacki},
  \citenamefont {Karpinski}, \citenamefont {Zhigadlo},\ and\ \citenamefont
  {Batlogg}}]{Moll2010}%
  \BibitemOpen
  \bibfield  {author} {\bibinfo {author} {\bibfnamefont {P.~J.~W.}\
  \bibnamefont {Moll}}, \bibinfo {author} {\bibfnamefont {R.}~\bibnamefont
  {Puzniak}}, \bibinfo {author} {\bibfnamefont {F.}~\bibnamefont {Balakirev}},
  \bibinfo {author} {\bibfnamefont {K.}~\bibnamefont {Rogacki}}, \bibinfo
  {author} {\bibfnamefont {J.}~\bibnamefont {Karpinski}}, \bibinfo {author}
  {\bibfnamefont {N.~D.}\ \bibnamefont {Zhigadlo}}, \ and\ \bibinfo {author}
  {\bibfnamefont {B.}~\bibnamefont {Batlogg}},\ }\href {\doibase
  doi:10.1038/nmat3489} {\bibfield  {journal} {\bibinfo  {journal} {Nature
  Materials}\ }\textbf {\bibinfo {volume} {9}},\ \bibinfo {pages} {628}
  (\bibinfo {year} {2010})}\BibitemShut {NoStop}%
\bibitem [{\citenamefont {Karpinski}\ \emph {et~al.}(2009)\citenamefont
  {Karpinski}, \citenamefont {Zhigadlo}, \citenamefont {Katrych}, \citenamefont
  {Bukowski}, \citenamefont {Moll}, \citenamefont {Weyeneth}, \citenamefont
  {Keller}, \citenamefont {Puzniak}, \citenamefont {Tortello}, \citenamefont
  {Daghero}, \citenamefont {Gonnelli}, \citenamefont {Maggio-Aprile},
  \citenamefont {Fasano}, \citenamefont {Fischer}, \citenamefont {Rogacki},\
  and\ \citenamefont {Batlogg}}]{Karpinski-PhysC-2009}%
  \BibitemOpen
  \bibfield  {author} {\bibinfo {author} {\bibfnamefont {J.}~\bibnamefont
  {Karpinski}}, \bibinfo {author} {\bibfnamefont {N.~D.}\ \bibnamefont
  {Zhigadlo}}, \bibinfo {author} {\bibfnamefont {S.}~\bibnamefont {Katrych}},
  \bibinfo {author} {\bibfnamefont {Z.}~\bibnamefont {Bukowski}}, \bibinfo
  {author} {\bibfnamefont {P.~J.~W.}\ \bibnamefont {Moll}}, \bibinfo {author}
  {\bibfnamefont {S.}~\bibnamefont {Weyeneth}}, \bibinfo {author}
  {\bibfnamefont {H.}~\bibnamefont {Keller}}, \bibinfo {author} {\bibfnamefont
  {R.}~\bibnamefont {Puzniak}}, \bibinfo {author} {\bibfnamefont
  {M.}~\bibnamefont {Tortello}}, \bibinfo {author} {\bibfnamefont
  {D.}~\bibnamefont {Daghero}}, \bibinfo {author} {\bibfnamefont
  {R.}~\bibnamefont {Gonnelli}}, \bibinfo {author} {\bibfnamefont
  {I.}~\bibnamefont {Maggio-Aprile}}, \bibinfo {author} {\bibfnamefont
  {Y.}~\bibnamefont {Fasano}}, \bibinfo {author} {\bibfnamefont
  {O.}~\bibnamefont {Fischer}}, \bibinfo {author} {\bibfnamefont
  {K.}~\bibnamefont {Rogacki}}, \ and\ \bibinfo {author} {\bibfnamefont
  {B.}~\bibnamefont {Batlogg}},\ }\href@noop {} {\bibfield  {journal} {\bibinfo
   {journal} {Physica C}\ }\textbf {\bibinfo {volume} {469}},\ \bibinfo {pages}
  {370} (\bibinfo {year} {2009})}\BibitemShut {NoStop}%
\bibitem [{\citenamefont {Weyeneth}\ \emph {et~al.}(2009)\citenamefont
  {Weyeneth}, \citenamefont {Puzniak}, \citenamefont {Mosele}, \citenamefont
  {Zhigadlo}, \citenamefont {Katrych}, \citenamefont {Bukowski}, \citenamefont
  {Karpinski}, \citenamefont {Kohout}, \citenamefont {Roos},\ and\
  \citenamefont {Keller}}]{Weyeneth-JSNM-2009}%
  \BibitemOpen
  \bibfield  {author} {\bibinfo {author} {\bibfnamefont {S.}~\bibnamefont
  {Weyeneth}}, \bibinfo {author} {\bibfnamefont {R.}~\bibnamefont {Puzniak}},
  \bibinfo {author} {\bibfnamefont {U.}~\bibnamefont {Mosele}}, \bibinfo
  {author} {\bibfnamefont {N.~D.}\ \bibnamefont {Zhigadlo}}, \bibinfo {author}
  {\bibfnamefont {S.}~\bibnamefont {Katrych}}, \bibinfo {author} {\bibfnamefont
  {Z.}~\bibnamefont {Bukowski}}, \bibinfo {author} {\bibfnamefont
  {J.}~\bibnamefont {Karpinski}}, \bibinfo {author} {\bibfnamefont
  {S.}~\bibnamefont {Kohout}}, \bibinfo {author} {\bibfnamefont
  {J.}~\bibnamefont {Roos}}, \ and\ \bibinfo {author} {\bibfnamefont
  {H.}~\bibnamefont {Keller}},\ }\href {\doibase 10.1007/s10948-008-0413-1}
  {\bibfield  {journal} {\bibinfo  {journal} {Journal of Superconductivity and
  Novel Magnetism}\ }\textbf {\bibinfo {volume} {22}},\ \bibinfo {pages} {325}
  (\bibinfo {year} {2009})}\BibitemShut {NoStop}%
\bibitem [{\citenamefont {Gordeev}\ \emph {et~al.}(2000)\citenamefont
  {Gordeev}, \citenamefont {Zhukov}, \citenamefont {de~Groot}, \citenamefont
  {Jansen}, \citenamefont {Gagnon},\ and\ \citenamefont
  {Taillefer}}]{Gordeev-PRL-2000}%
  \BibitemOpen
  \bibfield  {author} {\bibinfo {author} {\bibfnamefont {S.~N.}\ \bibnamefont
  {Gordeev}}, \bibinfo {author} {\bibfnamefont {A.~A.}\ \bibnamefont {Zhukov}},
  \bibinfo {author} {\bibfnamefont {P.~A.~J.}\ \bibnamefont {de~Groot}},
  \bibinfo {author} {\bibfnamefont {A.~G.~M.}\ \bibnamefont {Jansen}}, \bibinfo
  {author} {\bibfnamefont {R.}~\bibnamefont {Gagnon}}, \ and\ \bibinfo {author}
  {\bibfnamefont {L.}~\bibnamefont {Taillefer}},\ }\href@noop {} {\bibfield
  {journal} {\bibinfo  {journal} {Phys. Rev. Lett.}\ }\textbf {\bibinfo
  {volume} {85}},\ \bibinfo {pages} {4594} (\bibinfo {year}
  {2000})}\BibitemShut {NoStop}%
\bibitem [{\citenamefont {Baily}\ \emph {et~al.}(2008)\citenamefont {Baily},
  \citenamefont {Maiorov}, \citenamefont {Zhou}, \citenamefont {Balakirev},
  \citenamefont {Jaime}, \citenamefont {Foltyn},\ and\ \citenamefont
  {Civale}}]{Baily-PRL-2008}%
  \BibitemOpen
  \bibfield  {author} {\bibinfo {author} {\bibfnamefont {S.~A.}\ \bibnamefont
  {Baily}}, \bibinfo {author} {\bibfnamefont {B.}~\bibnamefont {Maiorov}},
  \bibinfo {author} {\bibfnamefont {H.}~\bibnamefont {Zhou}}, \bibinfo {author}
  {\bibfnamefont {F.~F.}\ \bibnamefont {Balakirev}}, \bibinfo {author}
  {\bibfnamefont {M.}~\bibnamefont {Jaime}}, \bibinfo {author} {\bibfnamefont
  {S.~R.}\ \bibnamefont {Foltyn}}, \ and\ \bibinfo {author} {\bibfnamefont
  {L.}~\bibnamefont {Civale}},\ }\href@noop {} {\bibfield  {journal} {\bibinfo
  {journal} {Phys. Rev. Lett.}\ }\textbf {\bibinfo {volume} {100}},\ \bibinfo
  {pages} {027004} (\bibinfo {year} {2008})}\BibitemShut {NoStop}%
\bibitem [{\citenamefont {Tokita}\ \emph {et~al.}(2009)\citenamefont {Tokita},
  \citenamefont {Nishizaki}, \citenamefont {Sasaki},\ and\ \citenamefont
  {Kobayashi}}]{Tokita-LowTemp-2009}%
  \BibitemOpen
  \bibfield  {author} {\bibinfo {author} {\bibfnamefont {Y.}~\bibnamefont
  {Tokita}}, \bibinfo {author} {\bibfnamefont {T.}~\bibnamefont {Nishizaki}},
  \bibinfo {author} {\bibfnamefont {T.}~\bibnamefont {Sasaki}}, \ and\ \bibinfo
  {author} {\bibfnamefont {N.}~\bibnamefont {Kobayashi}},\ }\href@noop {}
  {\bibfield  {journal} {\bibinfo  {journal} {JPCS}\ }\textbf {\bibinfo
  {volume} {150}},\ \bibinfo {pages} {052270} (\bibinfo {year}
  {2009})}\BibitemShut {NoStop}%
\bibitem [{\citenamefont {K\"upfer}\ \emph {et~al.}(2002)\citenamefont
  {K\"upfer}, \citenamefont {Ravikumar}, \citenamefont {Wolf}, \citenamefont
  {Zhukov}, \citenamefont {Will}, \citenamefont {Leibrock}, \citenamefont
  {Meier-Hirmer}, \citenamefont {W\"uhl},\ and\ \citenamefont
  {de~Groot}}]{Kupfer-PRB-2002}%
  \BibitemOpen
  \bibfield  {author} {\bibinfo {author} {\bibfnamefont {H.}~\bibnamefont
  {K\"upfer}}, \bibinfo {author} {\bibfnamefont {G.}~\bibnamefont {Ravikumar}},
  \bibinfo {author} {\bibfnamefont {T.}~\bibnamefont {Wolf}}, \bibinfo {author}
  {\bibfnamefont {A.~A.}\ \bibnamefont {Zhukov}}, \bibinfo {author}
  {\bibfnamefont {A.}~\bibnamefont {Will}}, \bibinfo {author} {\bibfnamefont
  {H.}~\bibnamefont {Leibrock}}, \bibinfo {author} {\bibfnamefont
  {R.}~\bibnamefont {Meier-Hirmer}}, \bibinfo {author} {\bibfnamefont
  {H.}~\bibnamefont {W\"uhl}}, \ and\ \bibinfo {author} {\bibfnamefont
  {P.~A.~J.}\ \bibnamefont {de~Groot}},\ }\href@noop {} {\bibfield  {journal}
  {\bibinfo  {journal} {Phys. Rev. B}\ }\textbf {\bibinfo {volume} {66}},\
  \bibinfo {pages} {064512} (\bibinfo {year} {2002})}\BibitemShut {NoStop}%
\bibitem [{\citenamefont {Malone}\ \emph {et~al.}(2009)\citenamefont {Malone},
  \citenamefont {Fletcher}, \citenamefont {Serafin}, \citenamefont
  {Carrington}, \citenamefont {Zhigadlo}, \citenamefont {Bukowski},
  \citenamefont {Katrych},\ and\ \citenamefont {Karpinski}}]{Malone-PRB-2009}%
  \BibitemOpen
  \bibfield  {author} {\bibinfo {author} {\bibfnamefont {L.}~\bibnamefont
  {Malone}}, \bibinfo {author} {\bibfnamefont {J.~D.}\ \bibnamefont
  {Fletcher}}, \bibinfo {author} {\bibfnamefont {A.}~\bibnamefont {Serafin}},
  \bibinfo {author} {\bibfnamefont {A.}~\bibnamefont {Carrington}}, \bibinfo
  {author} {\bibfnamefont {N.~D.}\ \bibnamefont {Zhigadlo}}, \bibinfo {author}
  {\bibfnamefont {Z.}~\bibnamefont {Bukowski}}, \bibinfo {author}
  {\bibfnamefont {S.}~\bibnamefont {Katrych}}, \ and\ \bibinfo {author}
  {\bibfnamefont {J.}~\bibnamefont {Karpinski}},\ }\href@noop {} {\bibfield
  {journal} {\bibinfo  {journal} {Phys. Rev. B}\ }\textbf {\bibinfo {volume}
  {79}},\ \bibinfo {pages} {140501 (R)} (\bibinfo {year} {2009})}\BibitemShut
  {NoStop}%
\bibitem [{\citenamefont {Koshelev}(2007)}]{Koshelev-PRB-2007}%
  \BibitemOpen
  \bibfield  {author} {\bibinfo {author} {\bibfnamefont {A.~E.}\ \bibnamefont
  {Koshelev}},\ }\href@noop {} {\bibfield  {journal} {\bibinfo  {journal}
  {Phys. Rev. B}\ }\textbf {\bibinfo {volume} {75}},\ \bibinfo {pages} {214513}
  (\bibinfo {year} {2007})}\BibitemShut {NoStop}%
\bibitem [{\citenamefont {Machida}(2003)}]{Machida-PRL-2003}%
  \BibitemOpen
  \bibfield  {author} {\bibinfo {author} {\bibfnamefont {M.}~\bibnamefont
  {Machida}},\ }\href@noop {} {\bibfield  {journal} {\bibinfo  {journal} {Phys.
  Rev. Lett.}\ }\textbf {\bibinfo {volume} {90}},\ \bibinfo {pages} {037001}
  (\bibinfo {year} {2003})}\BibitemShut {NoStop}%
\bibitem [{\citenamefont {Gropp}\ \emph {et~al.}(1996)\citenamefont {Gropp},
  \citenamefont {Kaper}, \citenamefont {Leaf}, \citenamefont {Levine},
  \citenamefont {Palumbo},\ and\ \citenamefont {Vinokur}}]{Gropp1996}%
  \BibitemOpen
  \bibfield  {author} {\bibinfo {author} {\bibfnamefont {W.~D.}\ \bibnamefont
  {Gropp}}, \bibinfo {author} {\bibfnamefont {H.~G.}\ \bibnamefont {Kaper}},
  \bibinfo {author} {\bibfnamefont {G.~K.}\ \bibnamefont {Leaf}}, \bibinfo
  {author} {\bibfnamefont {D.~M.}\ \bibnamefont {Levine}}, \bibinfo {author}
  {\bibfnamefont {M.}~\bibnamefont {Palumbo}}, \ and\ \bibinfo {author}
  {\bibfnamefont {V.~M.}\ \bibnamefont {Vinokur}},\ }\href@noop {} {\bibfield
  {journal} {\bibinfo  {journal} {J. Comput. Phys.}\ }\textbf {\bibinfo
  {volume} {123}},\ \bibinfo {pages} {256} (\bibinfo {year}
  {1996})}\BibitemShut {NoStop}%
\bibitem [{\citenamefont {Blatter}\ \emph {et~al.}(1994)\citenamefont
  {Blatter}, \citenamefont {Feigelman}, \citenamefont {Geshkenbein},
  \citenamefont {Larkin},\ and\ \citenamefont
  {Vinokur}}]{Blatter-RevModPhys-1994}%
  \BibitemOpen
  \bibfield  {author} {\bibinfo {author} {\bibfnamefont {G.}~\bibnamefont
  {Blatter}}, \bibinfo {author} {\bibfnamefont {M.~V.}\ \bibnamefont
  {Feigelman}}, \bibinfo {author} {\bibfnamefont {V.~B.}\ \bibnamefont
  {Geshkenbein}}, \bibinfo {author} {\bibfnamefont {A.~I.}\ \bibnamefont
  {Larkin}}, \ and\ \bibinfo {author} {\bibfnamefont {V.~M.}\ \bibnamefont
  {Vinokur}},\ }\href@noop {} {\bibfield  {journal} {\bibinfo  {journal} {Rev.
  Mod. Phys.}\ }\textbf {\bibinfo {volume} {66}},\ \bibinfo {pages} {1125}
  (\bibinfo {year} {1994})}\BibitemShut {NoStop}%
\bibitem [{\citenamefont {Baelus}\ \emph {et~al.}(2005)\citenamefont {Baelus},
  \citenamefont {Kadowaki},\ and\ \citenamefont {Peeters}}]{Baelus-PRB-2005}%
  \BibitemOpen
  \bibfield  {author} {\bibinfo {author} {\bibfnamefont {B.~J.}\ \bibnamefont
  {Baelus}}, \bibinfo {author} {\bibfnamefont {K.}~\bibnamefont {Kadowaki}}, \
  and\ \bibinfo {author} {\bibfnamefont {F.~M.}\ \bibnamefont {Peeters}},\
  }\href {\doibase 10.1103/PhysRevB.71.024514} {\bibfield  {journal} {\bibinfo
  {journal} {Phys. Rev. B}\ }\textbf {\bibinfo {volume} {71}},\ \bibinfo
  {pages} {024514} (\bibinfo {year} {2005})}\BibitemShut {NoStop}%
\bibitem [{\citenamefont {Choubey}\ \emph {et~al.}(2014)\citenamefont
  {Choubey}, \citenamefont {Berlijn}, \citenamefont {Kreisel}, \citenamefont
  {Cao},\ and\ \citenamefont {Hirschfeld}}]{Choubey-arxiv-2014}%
  \BibitemOpen
  \bibfield  {author} {\bibinfo {author} {\bibfnamefont {P.}~\bibnamefont
  {Choubey}}, \bibinfo {author} {\bibfnamefont {T.}~\bibnamefont {Berlijn}},
  \bibinfo {author} {\bibfnamefont {A.}~\bibnamefont {Kreisel}}, \bibinfo
  {author} {\bibfnamefont {C.}~\bibnamefont {Cao}}, \ and\ \bibinfo {author}
  {\bibfnamefont {P.~J.}\ \bibnamefont {Hirschfeld}},\ }\href@noop {}
  {\bibfield  {journal} {\bibinfo  {journal} {arXiv:1401.7732}\ } (\bibinfo
  {year} {2014})}\BibitemShut {NoStop}%
\bibitem [{\citenamefont {Oussena}\ \emph {et~al.}(1994)\citenamefont
  {Oussena}, \citenamefont {de~Groot}, \citenamefont {Gagnon},\ and\
  \citenamefont {Taillefer}}]{Oussena1994}%
  \BibitemOpen
  \bibfield  {author} {\bibinfo {author} {\bibfnamefont {M.}~\bibnamefont
  {Oussena}}, \bibinfo {author} {\bibfnamefont {P.~A.~J.}\ \bibnamefont
  {de~Groot}}, \bibinfo {author} {\bibfnamefont {R.}~\bibnamefont {Gagnon}}, \
  and\ \bibinfo {author} {\bibfnamefont {L.}~\bibnamefont {Taillefer}},\ }\href
  {\doibase 10.1103/PhysRevLett.72.3606} {\bibfield  {journal} {\bibinfo
  {journal} {Phys. Rev. Lett.}\ }\textbf {\bibinfo {volume} {72}},\ \bibinfo
  {pages} {3606} (\bibinfo {year} {1994})}\BibitemShut {NoStop}%
\bibitem [{\citenamefont {Lundqvist}\ \emph {et~al.}(2001)\citenamefont
  {Lundqvist}, \citenamefont {Rapp}, \citenamefont {Andersson},\ and\
  \citenamefont {Eltsev}}]{Lundquist-PRB-2001}%
  \BibitemOpen
  \bibfield  {author} {\bibinfo {author} {\bibfnamefont {B.}~\bibnamefont
  {Lundqvist}}, \bibinfo {author} {\bibfnamefont {O.}~\bibnamefont {Rapp}},
  \bibinfo {author} {\bibfnamefont {M.}~\bibnamefont {Andersson}}, \ and\
  \bibinfo {author} {\bibfnamefont {Y.}~\bibnamefont {Eltsev}},\ }\href@noop {}
  {\bibfield  {journal} {\bibinfo  {journal} {Phys. Rev. B}\ }\textbf {\bibinfo
  {volume} {64}},\ \bibinfo {pages} {060503(R)} (\bibinfo {year}
  {2001})}\BibitemShut {NoStop}%
\bibitem [{\citenamefont {Schnelle}\ \emph {et~al.}(1993)\citenamefont
  {Schnelle}, \citenamefont {Ernst},\ and\ \citenamefont
  {Wohlleben}}]{Schnelle-AnnPhys-1993}%
  \BibitemOpen
  \bibfield  {author} {\bibinfo {author} {\bibfnamefont {W.}~\bibnamefont
  {Schnelle}}, \bibinfo {author} {\bibfnamefont {P.}~\bibnamefont {Ernst}}, \
  and\ \bibinfo {author} {\bibfnamefont {D.}~\bibnamefont {Wohlleben}},\
  }\href@noop {} {\bibfield  {journal} {\bibinfo  {journal} {Ann. Phys.}\
  }\textbf {\bibinfo {volume} {2}},\ \bibinfo {pages} {109} (\bibinfo {year}
  {1993})}\BibitemShut {NoStop}%
\end{thebibliography}%

\end{document}